# Electronic properties of silicene in BN/silicene van der Waals heterostructures[*]

Ze-Bin Wu (吴泽宾), Yu-Yang Zhang (张余洋), Geng Li (李更),
Shixuan Du (杜世萱),[†] and Hong-Jun Gao (高鸿钧)

*Institute of Physics & University of Chinese Academy of Sciences, Chinese Academy of Sciences, Beijing 100190, China*
*Key Laboratory of Vacuum Physics, Chinese Academy of Sciences, Beijing 100190, China*

**Silicene is a promising 2D Dirac material as building block of van der Waals heterostructures (vdWHs). Here we investigate the electronic properties of hexagonal boron nitride/silicene (BN/Si) vdWHs using first-principles calculations. We calculate the energy band structures of BN/Si/BN heterostructure with different rotation angles and find the electronic properties of silicene is retained and protected robustly by BN layers. In BN/Si/BN/Si/BN heterostructure, we find the band structure near the Fermi energy is sensitive to the stacking configurations of the silicene layers due to interlayer coupling. The coupling is reduced by increasing the number of BN layers between silicene layers and become negligible in BN/Si/(BN)$_3$/Si/BN. In (BN)$_n$/Si superlattices, the band structure undergoes a conversion between Dirac lines and Dirac points with different number of BN layers between silicene layers. Calculations of silicene sandwiched by other 2D materials reveal that silicene sandwiched by low-carbon-doped boron nitride or HfO$_2$ is semiconducting.**

**Keywords:** silicene, BN, electronic property, heterostructure

[*] Project supported by the National Key Research and Development Program of China (No. 2016YFA0202300), the National Natural Science Foundation of China (Nos. 61390501, 61471337), the National Basic Research Program of China (No. 2013CBA01600), the CAS Pioneer Hundred Talents Program, and the Beijing Nova Program (No. Z181100006218023).
[†] Corresponding author. E-mail: sxdu@iphy.ac.cn

# 1. Introduction

Since the successful exfoliation of graphene in 2004,[1,2] more than a hundred intriguing 2D materials have been explored.[3-6] However, only a few of them are theoretically predicted to be Dirac materials.[7] The existence of Dirac cones in 2D materials gives rise to many novel physical properties, such as half-integer/fractional quantum Hall effect,[2,8-10] ultrahigh carrier mobility,[11] etc.[12] According to first-principles calculations,[13,14] monolayer silicene is a Dirac material with a stable buckled honeycomb structure.[15] The Fermi velocity of silicene is predicted to be ~$10^6$ m/s, and its intrinsic carrier mobility has the same order of magnitude as that of graphene.[16] Besides, the electronic properties of silicene have a better tunability than those of graphene.[17-22] These novel properties together with the possibly inherent compatibility with traditional silicon-based nanotechnology make silicene a promising candidate for electronic devices and a distinct building block for van der Waals layered materials.

Silicene has been grown on several substrates, such as Ag(111),[23-26] Ir(111),[27,28] Ru(0001),[29] and $ZrB_2$(0001).[30] Transistors have also been fabricated with epitaxial silicene on Ag(111) thin film.[31] However, there is no convincing experimental evidence supporting the existence of Dirac cones in epitaxial silicene because of the strong coupling with these substrates.[27,29,30,32-37] Silicene's surface is indeed very active because of the dangling Si bonds. The exposed silicene interacts easily with other active materials so that the electronic properties can be modified easily in unexpected ways. However, silicene has been found to interact weakly with graphene[38] and hexagonal boron nitride (BN)[17,39]. Dirac cones of silicene shift above the Fermi level when sandwiched by graphene. In contrast, BN does not cause the shift of Dirac cones. Therefore, BN-protected silicene could be a promising block for building van der Waals layered heterostructures (vdWHs). On the other hand, silicene sandwiched between some other 2D materials can be semiconducting, which is promising for applications in electronic devices.[40] Similar as the cases of molecules encapsulated by nanotubes or graphene layers, silicene is also energetically stabilized by the cladding layers.[41]

In this paper, we investigate electronic structures of BN-protected silicene in heterostructures and superlattices by first-principles calculations. Using "Si" to denote silicene, first we calculated the properties of BN/Si/BN heterostructures with different rotation angels and lateral shifts. Results show that rotation, lateral shift, and limited strain in silicene will not cause gap opening. To take advantage of the ultra-high carrier mobility of silicene, we propose a BN/Si/BN/Si/BN heterostructures, where each silicene layer is a conducting channel. We calculated the electronic properties of it and find that the band structure near the Fermi energy is sensitive to different stacking configurations of silicene layers due to interlayer coupling between silicene layers. We find that the coupling becomes negligible when silicene layers are separated by three BN layers in BN/Si/(BN)$_n$/Si/BN heterostructure. We also calculated potential (BN)$_n$/Si superlattices, and find that a conversion between Dirac points to Dirac lines can be achieved by changing the number of BN layers between silicene layers. Moreover, we calculated the band structures of silicene sandwiched by low-carbon-doped BN (l-BCN), high-carbon-doped BN (h-BCN), and HfO$_2$ and graphene/Si/BN. We find that the Dirac cones of silicene are opened in all of these heterostructures, but only l-BCN/Si/l-BCN and HfO$_2$/Si/HfO$_2$ are semiconducting with band gaps of 19 meV and 131 meV respectively.

## 2. Methods

All quantum-mechanical calculations were carried out based on the density functional theory (DFT) as implemented in the Vienna Ab-initio Simulation Package (VASP).[42-44] Electron-ion interactions were represented by projected augmented wave (PAW) potentials.[45] Exchange-correlation interactions were treated within the generalized gradient approximation (GGA-PBE).[46] The method of DFT-D3 to include van der Waals interactions was used in all the calculations of heterostructures.[47] The wave functions were expanded using a plane-wave basis set with an energy cutoff of 520 eV. For 2D structures, a vacuum layer larger than 20 Å was used to avoid interactions between layers in neighboring supercells. The force on each atom is relaxed to less than 0.01eV/Å. The Brillouin zone was sampled using a Γ-centered 18 × 18 mesh for (2 × 2) silicene in self-consistent calculations.

## 3. Results and discussions

## 3.1 Electronic properties of BN/Si/BN

It has been reported that the Dirac cone of silicene is protected when a (2 × 2) silicene is sandwiched by two (3 ×3) BN layers.[17] While in the real fabricating process, there are always lattice mismatches, rotation angles (misalignment), strains, and so on. To investigate the effects of above-mentioned situations in heterostructures, we build different BN/Si/BN supercells with different stacking configurations and calculate the energy band structures. The results are presented in table 1.

To eliminate the effect of strain in silicene in these heterostructures, the lattice constants of silicene is kept the same as that of its free-standing states and the heterostructures are otherwise fully relaxed. For all the BN/Si/BN heterostructures shown in Table 1, the calculated binding energy is 32 meV/Å$^2$. Compared with the binding energies for bilayer graphene which are 40 meV/Å$^2$ for A-A stacking and 42 meV/Å$^2$ for A-B stacking, the binding between BN and silicene is weaker. The binding energy is defined as $E_{Binding}$ = (2 × $E_{BN}$ + $E_{Si}$ – $E_{Heterostructure}$) / (2 × area) for BN/Si/BN and $E_{Binding}$ = (2 × $E_{Graphene}$ – $E_{Bilayer-graphene}$) / area for bilayer graphene. The negative $E_{Binding}$ value of BN/Si/BN indicates the stabilizing effect of the BN layers to silicene.

As a result of the small binding energy, there are probably different rotation angles in BN/Si heterostructures in the fabrication process in experiments. The calculations show that the electronic properties are nearly not affected by the rotation angles for BN/Si heterostructures due to weak van der Waals interaction. We also investigate the impact of lateral shifts in BN/Si/BN. Calculations show that lateral shifts almost have no effect on both the electronic properties and the total energy. Considering the possible strain in silicene layer induced by lattice mismatches in heterostructures, we calculate the energy band structure of silicene under strain of less than 3%. Calculation results show that the Dirac cones are not opened up and the electronic properties are nearly not affected by the strain. Moreover, the possible lattice distortion or rumpling induced by local strain would only have an effect on the local electronic properties. Therefore, we conclude that the electronic properties of silicene are protected robustly when sandwiched between BN layers.

## 3.2 Electronic properties of multilayer BN/Si/BN

Since silicene is left intact when sandwiched between BN layers under different rotation angles, lateral shifts, and limited strains in silicene, we use BN/Si/BN as a building block for multilayer heterostructures. To take advantage of the ultra-high carrier mobility of silicene, we propose a multilayer BN/Si heterostructures, where each silicene layer is a conducting channel. We first discuss potential BN/Si/BN/Si/BN heterostructures, in which two silicene layers are positioned in an A-A stacking configuration. The structures and calculation results of BN/Si/BN/Si/BN are shown in Fig. 1(a). As seen from the energy bands shown in Fig. 1(b), there are two sets of Dirac cones above and below the Fermi energy, marked by grey circles. The shift of these two sets of Dirac cones indicates that there is coupling between those two silicene layers. The energy separation of these two sets of Dirac cones is 143 meV. Because of the shift of Dirac cones from the Fermi level, there are crosses between the energy bands as shown in the energy band structures. As a result, a small gap of around 4 meV is opened due to perturbations at the Fermi level.

The distance between two silicene layers in BN/Si/BN/Si/BN is 7.6 Å. When two silicene layers are separated by a vacuum layer of 7.6 Å, there is nearly no interaction between them and the Dirac cones of them are overlapped. On the other hand, the energy states of BN/Si/BN/Si/BN near the Fermi level ($E_f \pm 0.3$ eV) are contributed by both silicene and BN layers, as seen from the partial charge density in Fig. 1(c). Therefore, we conclude that the interlayer coupling between silicene layers is mediated by the BN layer between them. According to this result, a question arises that whether more BN layers between silicene layers decrease the coupling or not.

To further investigate the interlayer coupling and its impact on the electronic properties, we increase the number of BN layers between silicene layers. The energy band structures of the BN/Si/(BN)$_n$/Si/BN and BN/Si/(BN)$_n$/Si/BN structures are shown in Figs. 1(d) and 1(e), respectively. As seen in Fig. 1(d), the energy difference between the two sets of Dirac cones reduces to 18 meV, indicating that the interlayer coupling is weakened by increasing the number of BN layers. The shift of the Dirac cones becomes negligible when the two silicene layers are separated by three BN layers, as shown in Fig. 1(e). In short, the interaction between silicene layers is weakened by increasing the number of BN layers. Therefore, the proposed multi-conducting-channel BN/Si heterostructure is only possible when silicene layers are separated by three or more than three BN layers.

In the heterostructures mentioned above, only A-A stacking silicene layers are considered. Because of the weak binding between BN and silicene in BN/Si heterostructures, there can be different sequences for silicene layers during fabrication in experiments. We investigate the impact of different stacking configurations of those two silicene layers resembling bilayer graphene. For graphene, there are two common stacking configurations, A-A stacking and A-B stacking. Different from graphene, there are five different high-symmetry stacking configurations for two silicene layers due to its buckled structure. The heterostructures and the energy band structures near the Fermi level are shown in Fig. 2. The results of A-A stacking configurations are shown in Fig. 2(a) and 2(b) and the results of A-B stacking configuration are shown in Fig. 2(c)-(e). The energy band structure of an A-B stacking configuration is close to each other whereby the Dirac cones are destroyed, with a negligible gap appearing at K points. In an A-A stacking configuration, the Dirac cones are preserved but shifted to different sides of the Fermi level as discussed before. For A-A stacking configuration in mirror symmetry, the Dirac cones are destroyed. The total binding energy of BN/Si/BN/Si/BN is almost the same of around 116 meV/Å$^2$. (The binding energy are calculated as $E_{Binding}$ = (3 × $E_{BN}$ + 2 × $E_{Si}$ − $E_{Heterostructure}$) / area.)

### 3.3 Electronic properties of BN/Si superlattices

We further investigated $(BN)_n$/silicene superlattices, in which silicene layers are positioned in A-A configuration, with n=1, 2, 3. The unit cell and the first Brillouin zone are shown in Fig. 3(a). Band structures of the superlattices are shown in Fig. 3(b)-(d). Consistent with the results above, the Dirac cones shift from the Fermi level when silicene layers are separated by less than three BN layers, forming Dirac lines along the K-H lines in the reciprocal space as seen in Fig 3(b) and 3(c). When the number of BN layers is increased to three, the Dirac cones return to the positions of $E_f \pm 2$ meV, indicating that the coupling is negligible between silicene layers and quasi-free-standing silicene is achieved in this van der Waals superlattice. T. P. Kaloni et al[48] reported that quasi-free-standing silicene can be achieved in Si/BN superlattice where adjacent silicene layers are separated by only one layer of BN. However, according to our result, one layer of BN is not enough to isolate silicene layers in BN/Si superlattice and there are Dirac lines but not Dirac points. Based on the results, it is safer to use at least three BN layers as cladding layers at each side of silicene to protect the electronic properties from being modified.

## 3.4 Semiconducting silicene in heterostructures

It is important to open a gap in silicene for its application in electronic devices. Many different ways based on theoretical calculations have been reported to achieve this goal.[17-22] Here we present the results that silicene sandwiched by l-BCN or $HfO_2$ is semiconducting. We calculated the band structures of Gr/Si/BN, BCN/Si/BCN, and $HfO_2$/Si/HfO2 heterostructures. BCN can be fabricated from either BN[49] or graphene[50] and has a tunable gap depending on the doping concentration. The atomic models of l-BCN and h-BCN used in our calculations are presented in Fig. 4(a). Based on the calculated phonon dispersions, there are no virtual frequencies for both l-BCN and h-BCN, which indicates that they are stable materials.

Dirac cones of silicene in Gr/Si/BN are slightly opened because of the breaking of the space-inversion, as shown in Fig. 4(b). The Fermi level lies below the Dirac cones due to the presence of graphene.[38] In both l-BCN/Si/l-BCN and h-BCN/Si/h-BCN, the Dirac cones are opened, with energy gaps of 19 meV and 28 meV. In the absence of doping at T = 0 K, the Fermi level lies in the middle of the gap in l-BCN/Si/l-BCN (Fig. 1(e)). While in h-BCN/Si/h-BCN, the Fermi level is below the opened Dirac cones (Fig. 1(f)).

$HfO_2$ is a large-gap semiconductor, like BN. However, in $HfO_2$/Si/HfO2 (Fig. 4(e)), a gap of 131 meV is opened as shown in Fig. 4(f). This gap may be caused by the interaction between silicon and oxygen. Therefore, $HfO_2$/Si/HfO2 are semiconducting. The binding energy of $HfO_2$/Si/HfO2 is 45 meV/Å$^2$ which is larger than that of BN/Si/BN, indicating the $HfO_2$ layers have a stronger stabilization effect to silicene than BN.

## 4. Conclusion

In BN/Si/BN heterostructure with different rotation angles, lateral shifts, and limited strains, the electronic property of silicene is protected robustly. In BN/Si/BN/Si/BN heterostructure, the band structure is sensitive to the stacking configurations of the silicene layers due to interlayer coupling. The coupling can be inhibited by increasing the number of BN layers and becomes negligible at three BN layers. In BN/Si superlattices with silicene positioned in A-A stacking configuration, a conversion between Dirac lines and Dirac points can be achieved by changing the number of BN layers between silicene layers. Therefore, quasi-free-standing silicene can only be achieved in BN/Si superlattice when adjacent silicene layers are separated by at least three BN layers. Multi-conducting-channel BN/Si heterostructures is only possible when silicene layers are separated by no less than three BN layers. Besides, it is safer to use at least three BN layers on each side of silicene to protect the Dirac cones. At last, silicene sandwiched by l-BCN or $HfO_2$ is semiconducting.

## Acknowledgements

We thank Prof. Sokrates T. Pantelides in Vanderbilt University for providing advice on this project. Thanks to National Supercomputing Center in Tianjin for providing computational resources.

**Table 1.** Silicene sandwiched by BN with different rotation angles and lattice mismatches.

| BN/Si/BN | | Rotation angle | Lattice mismatch[1] | Binding energy meV/ Å² | Dirac cones preserved (Y/N)[2] |
| --- | --- | --- | --- | --- | --- |
| Supercell | Top view | | | | |
| 3BN/2Si/3BN | 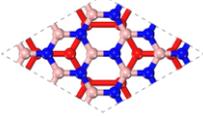 | 0° | 2.3% | 32 | Y |
| √31BN/√13Si/√31BN | 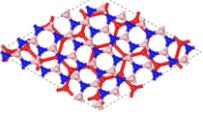 | 5° | −0.6% | 34 | Y |
| √7BN/√3Si/√7BN | 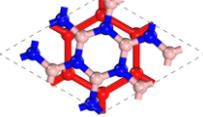 | 10.9° | 0.5% | 33 | Y |
| √39BN/4Si/√39BN | 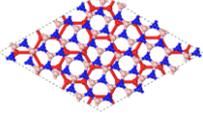 | 16.1° | -1.7% | 35 | Y |
| 4BN/√7Si/4BN | 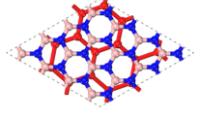 | 19.1° | 1.5% | 32 | Y |

[1] + (-) indicates tensile (compressive) strain in BN layers.

[2] Y (N) indicates the Dirac cones of silicene are retained (destroyed).

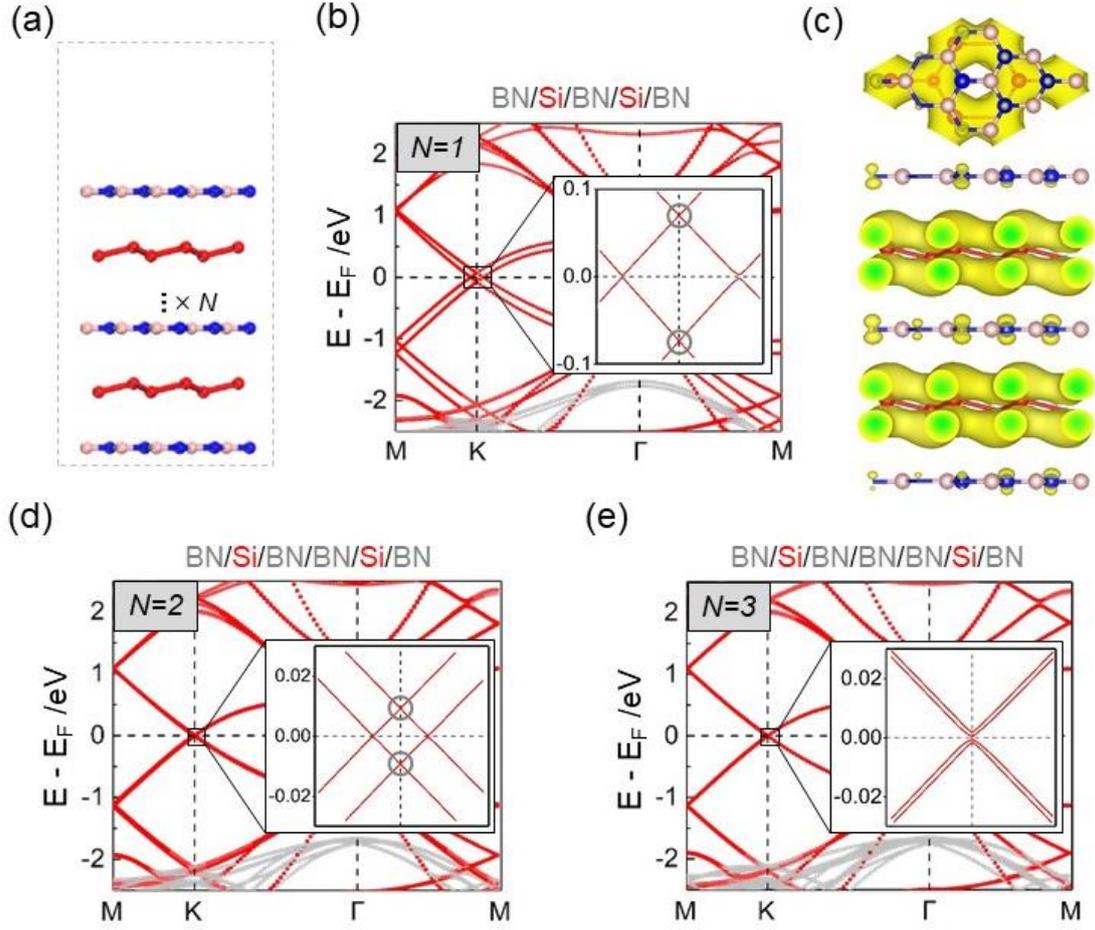

**Fig. 1.** (color online) Coupling between silicene layers intercalated by BN layers. (a): Supercell of a 2D vdWHs with two silicene layers encapsulated inside but separated by $N$ BN layers. (b): Energy band structure of supercell shown in (a) with $N = 1$. (c): Partial charge density near the Fermi level ($E_f \pm 0.3$ eV) related to band structure shown in b. (d)-(e): Energy band structures of the supercell shown in (a) with $N = 2$ and 3, respectively.

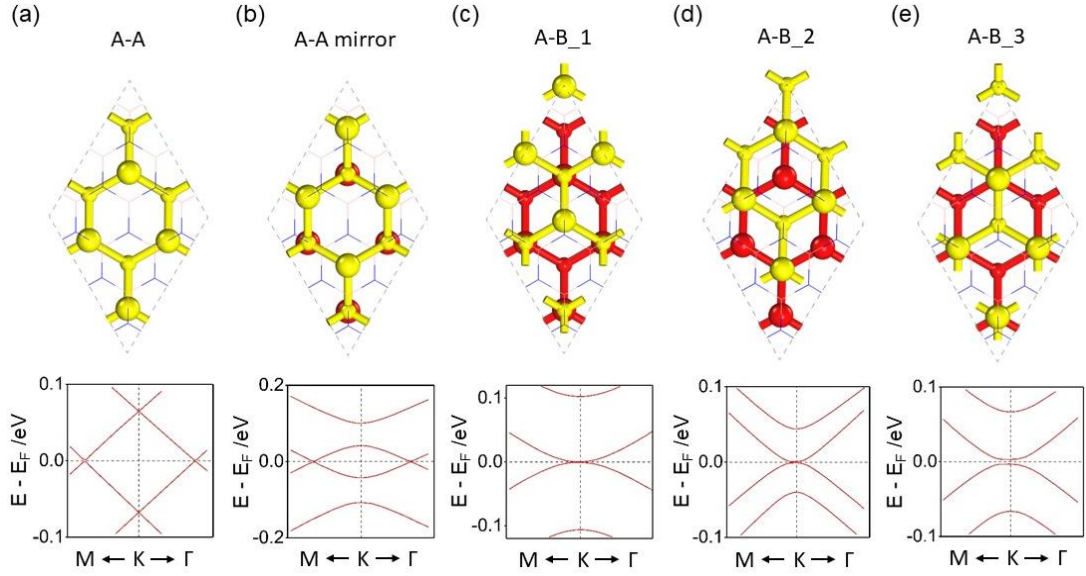

**Fig. 2.** (color online) Top views and energy band structures of BN/Si/BN/Si/BN with different stacking configurations of silicene layers. To highlight the silicene layers, BN layers are shown in line style and the two silicene layers are colored by red and yellow respectively. The silicene layer colored in red is kept still and below. Two sub-lattices of silicene are indicated by the sizes of the balls that the larger balls represent the higher atoms in this top view. (a)-(b): The band structure of A-A stacking configuration. (c)-(e): The band structure of A-B stacking configuration.

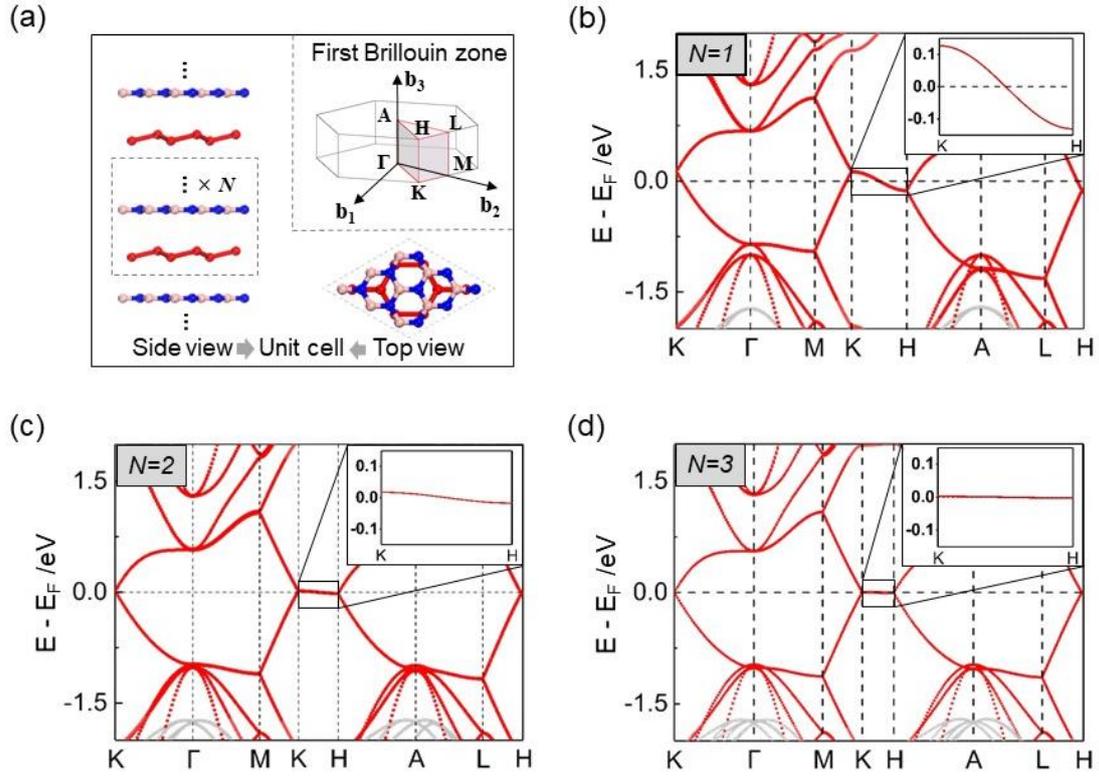

**Fig. 3.** (color online) Quasi-free-standing silicene in superlattices. (a): Unit cell and first Brillouin zone of the BN/silicene superlattices. (b)-(d): Energy band structures of the superlattices, differing from each other by the number of BN layers between adjacent silicene layers, from one to three.

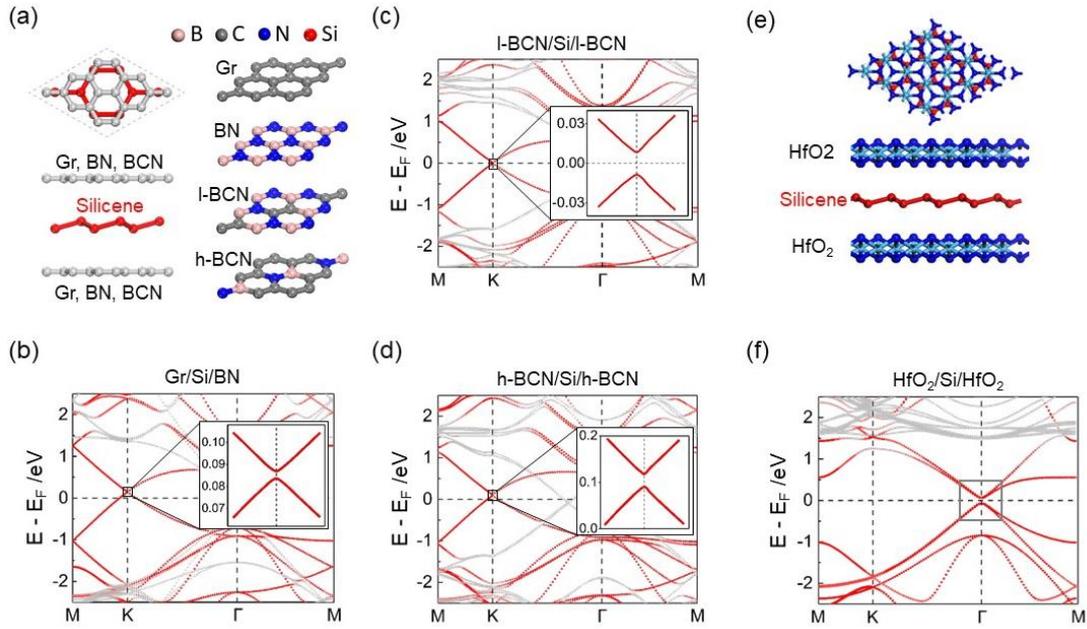

**Fig. 4.** (color online) Silicene sandwiched by other 2D materials where the Dirac cones are opened. (a): Left panel: Sandwich heterostructure with silicene in the middle. Right panel: Unit cells of cladding layers. (b)-(d): Projected energy band structures of Gr/Si/BN, BCN/Si/BCN sandwich heterostructures. Contributions of silicene and cladding layers are quantitatively indicated by red-grey color scale, where a 100% contribution from silicene is marked by pure red. (e): Silicene sandwiched by $HfO_2$. (f): The projected energy band structure of $HfO_2/Si/HfO_2$.

# References

[1] Novoselov K S, Geim A K, Morozov S V, Jiang D, Zhang Y, Dubonos S V, Grigorieva I V, and Firsov A A 2004 *Science* **306** 666
[2] Zhang Y, Tan Y-W, Stormer H L, and Kim P 2005 *Nature* **438** 201
[3] Xu M, Liang T, Shi M, and Chen H 2013 *Chem. Rev.* **113** 3766
[4] Butler S Z, Hollen S M, Cao L, Cui Y, Gupta J A, Gutiérrez H R, Heinz T F, Hong S S, Huang J, Ismach A F, Johnston-Halperin E, Kuno M, Plashnitsa V V, Robinson R D, Ruoff R S, Salahuddin S, Shan J, Shi L, Spencer M G, Terrones M, Windl W, and Goldberger J E 2013 *ACS Nano* **7** 2898
[5] Bhimanapati G R, Lin Z, Meunier V, Jung Y, Cha J, Das S, Xiao D, Son Y, Strano M S, Cooper V R, Liang L, Louie S G, Ringe E, Zhou W, Kim S S, Naik R R, Sumpter B G, Terrones H, Xia F, Wang Y, Zhu J, Akinwande D, Alem N, Schuller J A, Schaak R E, Terrones M, and Robinson J A 2015 *ACS Nano* **9** 11509
[6] Naguib M, Mochalin V N, Barsoum M W, and Gogotsi Y 2014 *Adv. Mater.* **26** 992
[7] Wang J, Deng S, Liu Z, and Liu Z 2015 *Natl. Sci. Rev.* **2** 22
[8] Novoselov K S, Geim A K, Morozov S V, Jiang D, Katsnelson M I, Grigorieva I V, Dubonos S V, and Firsov A A 2005 *Nature* **438** 197
[9] Bolotin K I, Ghahari F, Shulman M D, Stormer H L, and Kim P 2009 *Nature* **462** 196
[10] Du X, Skachko I, Duerr F, Luican A, and Andrei E Y 2009 *Nature* **462** 192
[11] Bolotin K I, Sikes K J, Jiang Z, Klima M, Fudenberg G, Hone J, Kim P, and Stormer H L 2008 *Solid State Commun.* **146** 351
[12] Castro Neto A H, Guinea F, Peres N M R, Novoselov K S, and Geim A K 2009 *Rev. Mod. Phys.* **81** 109
[13] Cahangirov S, Topsakal M, Aktürk E, Şahin H, and Ciraci S 2009 *Phys. Rev. Lett.* **102** 236804
[14] Şahin H, Cahangirov S, Topsakal M, Bekaroglu E, Akturk E, Senger R T, and Ciraci S 2009 *Phys. Rev. B* **80** 155453
[15] Li H, Hui-Xia F, and Meng S 2015 *Chin.Phys.B* **24** 086102
[16] Shao Z-G, Ye X-S, Yang L, and Wang C-L 2013 *J. Appl. Phys.* **114** 093712
[17] Ni Z, Liu Q, Tang K, Zheng J, Zhou J, Qin R, Gao Z, Yu D, and Lu J 2012 *Nano Lett.* **12** 113
[18] Scalise E, Houssa M, Cinquanta E, Grazianetti C, Broek B v d, Pourtois G, Stesmans A, Fanciulli M, and Molle A 2014 *2D Materials* **1** 011010
[19] Gao N, Li J C, and Jiang Q 2014 *Phys. Chem. Chem. Phys.* **16** 11673
[20] Du Y, Zhuang J, Liu H, Xu X, Eilers S, Wu K, Cheng P, Zhao J, Pi X, See K W, Peleckis G, Wang X, and Dou S X 2014 *ACS Nano* **8** 10019
[21] Quhe R, Fei R, Liu Q, Zheng J, Li H, Xu C, Ni Z, Wang Y, Yu D, Gao Z, and Lu J 2012 *Sci. Rep.* **2** 853
[22] Drummond N D, Zólyomi V, and Fal'ko V I 2012 *Phys. Rev. B* **85** 075423
[23] Chun-Liang L, Ryuichi A, Kazuaki K, Noriyuki T, Emi M, Yousoo K, Noriaki T, and Maki K 2012 *Appl. Phys. Express* **5** 045802
[24] Feng B, Ding Z, Meng S, Yao Y, He X, Cheng P, Chen L, and Wu K 2012 *Nano Lett.* **12** 3507
[25] Chen L, Liu C-C, Feng B, He X, Cheng P, Ding Z, Meng S, Yao Y, and Wu K 2012 *Phys. Rev. Lett.* **109** 056804
[26] Vogt P, De Padova P, Quaresima C, Avila J, Frantzeskakis E, Asensio M C,

Resta A, Ealet B, and Le Lay G 2012 *Phys. Rev. Lett.* **108** 155501

[27]  Meng L, Wang Y, Zhang L, Du S, Wu R, Li L, Zhang Y, Li G, Zhou H, Hofer W A, and Gao H-J 2013 *Nano Lett.* **13** 685
[28]  Meng L, Wang Y-L, Zhang L-Z, Du S-X, and Gao H-J 2015 *Chin.Phys.B* **24** 086803
[29]  Huang L, Zhang Y-F, Zhang Y-Y, Xu W, Que Y, Li E, Pan J-B, Wang Y-L, Liu Y, Du S-X, Pantelides S T, and Gao H-J 2017 *Nano Lett.* **17** 1161
[30]  Fleurence A, Friedlein R, Ozaki T, Kawai H, Wang Y, and Yamada-Takamura Y 2012 *Phys. Rev. Lett.* **108** 245501
[31]  Tao L, Cinquanta E, Chiappe D, Grazianetti C, Fanciulli M, Dubey M, Molle A, and Akinwande D 2015 *Nature Nano.* **10** 227
[32]  Guo Z-X, Furuya S, Iwata J-i, and Oshiyama A 2013 *Phys. Rev. B* **87** 235435
[33]  Wang Y-P and Cheng H-P 2013 *Phys. Rev. B* **87** 245430
[34]  Arafune R, Lin C L, Nagao R, Kawai M, and Takagi N 2013 *Phys. Rev. Lett.* **110** 229701
[35]  Lin C-L, Arafune R, Kawahara K, Kanno M, Tsukahara N, Minamitani E, Kim Y, Kawai M, and Takagi N 2013 *Phys. Rev. Lett.* **110** 076801
[36]  Cahangirov S, Audiffred M, Tang P, Iacomino A, Duan W, Merino G, and Rubio A 2013 *Phys. Rev. B* **88** 035432
[37]  Zhong H-X, Quhe R-G, Wang Y-Y, Shi J-J, and Lü J 2015 *Chin.Phys.B* **24** 087308
[38]  Neek-Amal M, Sadeghi A, Berdiyorov G R, and Peeters F M 2013 *Appl. Phys. Lett.* **103** 261904
[39]  Wang M, Liu L, Liu C-C, and Yao Y 2016 *Phys. Rev. B* **93** 155412
[40]  Kou L, Ma Y, Yan B, Tan X, Chen C, and Smith S C 2015 *ACS Appl. Mater. Interfaces* **7** 19226
[41]  Smeu M, Zahid F, Ji W, Guo H, Jaidann M, and Abou-Rachid H 2011 *J. Phys. Chem. C* **115** 10985
[42]  Kresse G and Furthmüller J 1996 *Comp. Mat. Sci.* **6** 15
[43]  Kresse G and Furthmüller J 1996 *Phys. Rev. B* **54** 11169
[44]  Kresse G and Hafner J 1993 *Phys. Rev. B* **47** 558
[45]  Kresse G and Joubert D 1999 *Phys. Rev. B* **59** 1758
[46]  Perdew J P, Burke K, and Ernzerhof M 1996 *Phys. Rev. Lett.* **77** 3865
[47]  Grimme S, Antony J, Ehrlich S, and Krieg H 2010 *J. Chem. Phys.* **132** 154104
[48]  Kaloni T P, Tahir M, and Schwingenschlögl U 2013 *Sci. Rep.* **3** 3192
[49]  Huang C, Chen C, Zhang M, Lin L, Ye X, Lin S, Antonietti M, and Wang X 2015 *Nature Communi.* **6** 7698
[50]  Ba K, Jiang W, Cheng J, Bao J, Xuan N, Sun Y, Liu B, Xie A, Wu S, and Sun Z 2017 *Sci. Rep.* **7** 45584